\documentclass{PoS}
\renewcommand{\deg}{$^{\circ}$}
\newcommand{\citep}{\cite}
\newcommand{\citet}{\cite}
\newcommand{\MV}{MultiView}
\newcommand{\uas}{$\mu$as}
\newcommand{\amin}{$^\prime$}

\title{(Ultra) Precise Astrometry today and tomorrow, with Next-generation Observatories.}

\ShortTitle{Microarcsecod Astrometry with Next-generation Observatories}

\author{\speaker{Mar\'ia Rioja}\\
        ICRAR/UWA,  CASS/CSIRO (Australia), OAN/IGN (Spain)\\
        E-mail: \email{maria.rioja@icrar.org}\\
        {Richard Dodson}\\
        ICRAR/UWA (Australia)\\
        E-mail: \email{richard.dodson@icrar.org}
        }

\abstract{High precision astrometry provides the foundation to resolve many fundamental problems in astrophysics. The application of astrometric studies spans a wide range of fields, and has undergone
enormous growth in recent years. This is as a consequence of the increasing measurement precision and wide applicability, which is due in turn to the development of new techniques.
Forthcoming next generation observatories have the potential to further increase the astrometric precision, providing there is a matching improvement in the methods to correct for systematic errors.
The EVN and other observatories are providing demonstrations of these and are acting as pathfinders for next-generation telescopes such as the SKA and ngVLA.
We will review the perspectives for the coming facilities and examples of the current state-of-the-art for astrometry.}

\FullConference{14th European VLBI Network Symposium \& Users Meeting (EVN 2018)\\
		8-11 October 2018\\
		Granada, Spain}

\begin{document}

\section{Introduction}
Astrometry provides the foundation to attack many problems in astrophysics, and VLBI has the potential to achieve the largest angular resolutions and highest astrometric precision in astronomy. 
The astrometric potential is ultimately determined by the length of the baselines, the relentless improvements to increase the sensitivity and the observing frequency,
\underline{provided} that the systematic errors can be accounted for. 
%
%
These are dominated by atmospheric propagation effects that have the largest impact on observational astrometric performance, and astrometry depends on its mitigation using suitable calibration techniques. 
Conventional Phase Referencing (PR) derives the atmospheric corrections for the target from contemporaneous (i.e. in-beam) or alternating observations of a calibrator source and is most applicable at cm-wavelengths.
The astrometric errors are strongly dependent on the angular separation, switching time and the observing frequency. 

The quest for high precision astrometry has driven the developments of advanced calibration methods to be used with existing VLBI networks. Huge strides have been made in the last 15 years that make it possible to achieve ca. 10\, microarcsecond (\uas\ ) astrometric precision in a nearly regular fashion in a range of moderate frequencies dominated by tropospheric errors; from 8\,GHz to 22\,GHz, and up to a maximum of 43\,GHz in a limited number of cases. 
The implementation of advanced tropospheric calibration (ATC {\em hereafter}) methods, such as ``geodetic-blocks'' and others \cite{reid_micro,honma_08}, combined with PR techniques, allowed precise astrometry with calibrators up to 1\deg--2\deg\ away.
The wide applicability and astrometric performance has led to an exponential increase in the fields of research accessible to astrometry, as described in \cite{reid_micro}. 
Nevertheless, the application outside of this moderate frequency range remained limited or non-existent.

More recently, routine high precision astrometry up to 130\,GHz has become possible using mm-VLBI observations with the Korean VLBI Network (KVN) \cite{sslee_14}, combined with the Source Frequency Phase Referencing (SFPR) calibration methods \cite{rioja_11a}. This breakthrough has come from parallel developments on technological solutions and calibration methods. 
The KVN is the first dedicated mm-VLBI  network, and is equipped with an innovative simultaneous multi-frequency receiver system \cite{han_08,han_13}, for a precise atmospheric compensation. 
These developments are relevant for VLBI observations in the ALMA- and ngVLA-eras.

In comparison, developments for the low frequency regime ($\nu\lesssim$8\,GHz), dominated by a different type of errors arising from the ionospheric propagation, have lagged behind.
However in the last years this domain has had a renaissance, driven by the SKA project.
Achieving the SKA science goals also requires a parallel development of advanced calibration methods, along with the technological solutions.
%
In this case, our proposed strategy to achieve ultra precise astrometry on regular basis is to combine large collecting areas with MultiBeam technology and \MV\ (MV) calibration \cite{rioja_17}, a method that we have been developing over the last few years. 

In this paper we describe the technological solutions and calibration methods that enable $\mu$as astrometry across the radio spectrum, at high frequencies (Sect. 2) and at low frequencies (Sect. 3), a limited review of the current state-of-the-art astrometrical results since the last EVN Symposium (Sect. 4) and ongoing technological developments relevant to astrometry (Sect. 5).

\section{mm-VLBI Microarcsecond astrometry using multi-frequency methods}

\subsection{Troposphere and conventional PR}

The tropospheric propagation imposes fast phase fluctuations onto the incoming radio-waves and are the dominant error contribution in observations at $\nu \gtrsim$8\,GHz. 
The effect increases proportionally to the observing frequency and has a direct impact on the coherence times; this, and the fact that the sources are intrinsically weaker at higher frequencies and the instrumental noise is higher, results in a scarcity of suitable calibrator sources that has limited the application of PR and ATC methods to 8--43\,GHz. 
Hence traditionally there has been a distinct lack of astrometric measurements at $\nu>$43\,GHz  (other than a unique case at 86\,GHz \cite{porcas_02}). 
The increased sensitivity from VLBI observations with next-generation instruments such as the ngVLA will alleviate this situation, given the improved performance and increased collecting area.
Nevertheless the quest for precision astrometry at higher frequencies will continue to pose a challenge and it is worth exploring innovative methods for very precise tropospheric compensation for mm-VLBI astrometry. 


\subsection{SFPR/MFPR astrometric methods and demonstration with existing instruments}

An approach that has only recently begun to deliver on its promise for mm-VLBI astrometry is Multi-Frequency Calibration, based on the non-dispersive nature of the tropospheric propagation effects. 
That is, for example, using dual frequency observations of the target source, so that the lower (and more amenable)  frequency can be used to compensate
for tropospheric residuals in the high frequency observations.
Such strategies have been proposed in the past \cite{asaki_fpt_96,carilli_99,middelberg_05}, but it was not until the development of calibration methods that precisely accounted for all, non-dispersive and dispersive, error contributions that successful VLBI astrometry was achieved (i.e. SFPR and Multi-Frequency Phase Referencing (MFPR)).
The basis of the SFPR method, along with a comprehensive astrometric error analysis and observational demonstration has been presented elsewhere \cite{rioja_11a,rioja_14}.
The dispersive errors are mitigated with interleaving observations of a second source, with a duty cycle up to many minutes, and angular separation up to many degrees. 
MFPR is an alternative method that explicitly measures the dispersive terms on the target source itself, using additional observations between 1.3 and 22-GHz and is fully described in \cite{dodson_17}.
%

The first demonstration of their astrometric capability was carried out using fast frequency switching VLBA observations, in 2007. 
Superior performance comes from the technical solution adopted by the KVN, the multi-frequency receivers \cite{han_08,han_13} that enable simultaneous observations at four frequency bands (22, 43, 86, 129\,GHz). This results in much improved error mitigation and therefore allows applicability to much higher frequencies. Rioja \cite{rioja_15} describes a SFPR demonstration at the highest KVN band, at 130\,GHz (see Figure 1).

\begin{figure}
    \centering
    \includegraphics[width=\textwidth]{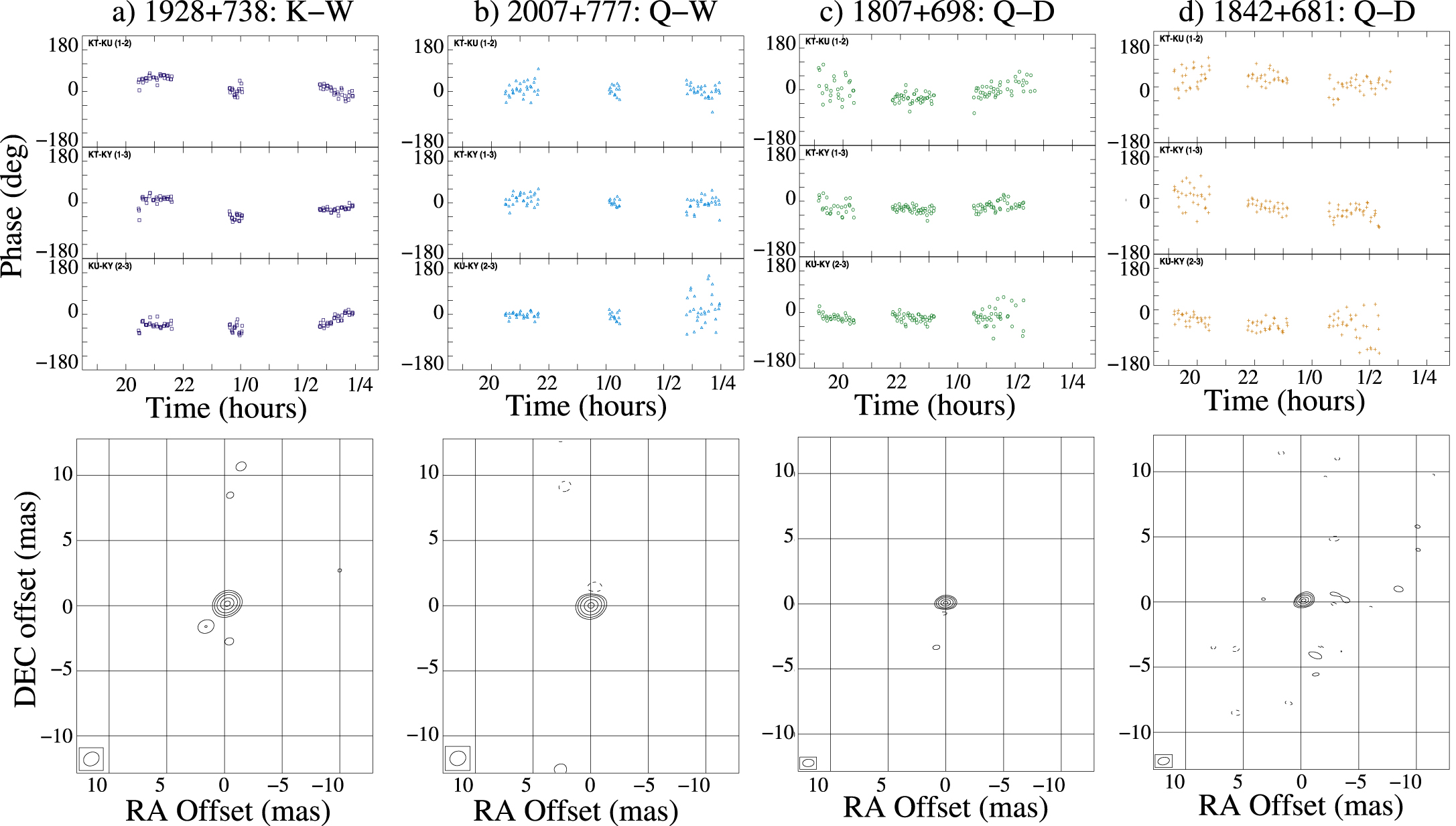}
    \caption{
        Outcomes of SFPR astrometric analysis using KVN observations at the four frequency bands K/Q/W/D (22/43/86/130 GHz, respectively),  from \cite{rioja_15}. Upper row: SFPRed residual visibility phases for source and frequency pairs. The target source and target frequency band (i.e., ${\nu }^{{\rm{high}}}$), along with the angular separation from reference source, the reference frequency band (i.e., ${\nu }^{{\rm{low}}}$), and the frequency ratio R, in parentheses, are specified next  for each plot. From left to right: 1928+738 at W band (6\deg\ apart on the sky, K band, R = 4); 2007+777 at W band (6\deg\ apart, Q band, R = 2); 1807+698 at D band (8\deg\ apart, Q band, R = 3); 1842+681 at D band (10\deg\ apart, Q band, R = 3). Lower row: SFPRed astrometric maps from left to right: 1928+738 at 87\,GHz (W band), 2007+777 at 87\,GHz; 1807+698 and 1842+681 at 130\,GHz (D band). Peak fluxes are 2 Jy beam$^{-1}$, 266 mJy beam$^{-1}$, 415 mJy beam$^{-1}$, and 216 mJy beam$^{-1}$, respectively. The contour levels in the maps start from 0.75\% of the corresponding peak fluxes, respectively, and double thereafter in all cases. Each map includes a negative contour level at the same percent level of the peak flux as the first positive one. The beam size is indicated at the bottom left of the image.}
    \label{fig:sfpr130}
\end{figure}



With an increasing number of telescopes with multi-frequency facilities the perspectives for mm-VLBI are most promising.
The KVN is leading the effort towards exploring the advantages of this technical solution at the highest frequencies, and together with the EAVN, for longer baselines.
%
In addition, in Europe, Yebes-40m has a multi-frequency system installed and carries out regular observations with KVN and KaVA \cite[these proc.]{bws_evn_18};
several other telescopes are expected to be simultaneous multi-frequency ready in the near future, both in Europe and across the globe.
%
Also the specifications of ngVLA include the (near)simultaneous multi-frequency capability, and is further discussed in Section \ref{sec:ngvla}.
We expect that the current promising results will further improve with the ongoing developments and observations with the next-generation instruments. 

\subsection{Astrometry in the ngVLA era}
\label{sec:ngvla}

The ngVLA is an American proposal to address the science goals considered for SKA-High \cite{ngvla}. It will have long baselines, spanning at least $\sim$500\,km, and cover the frequency range from 1.2 to 120\,GHz \cite{ngvla_sci}, with a similar collecting area to SKA-1. The technological solutions are challenging \cite{ngvla_tec}, and we have been using KVN as a pathfinder, as it has a similar maximum frequency and baseline length. The ngVLA multi-frequency capabilities are not yet finalised, but will either be simultaneous or extremely fast frequency switching. Both options are SFPR compatible, and combined with the huge sensitivity would allow ultra precise \uas-astrometry on regular basis, to many targets, with mm-VLBI.

\begin{figure}
    \centering
    \includegraphics[width=0.6\textwidth]{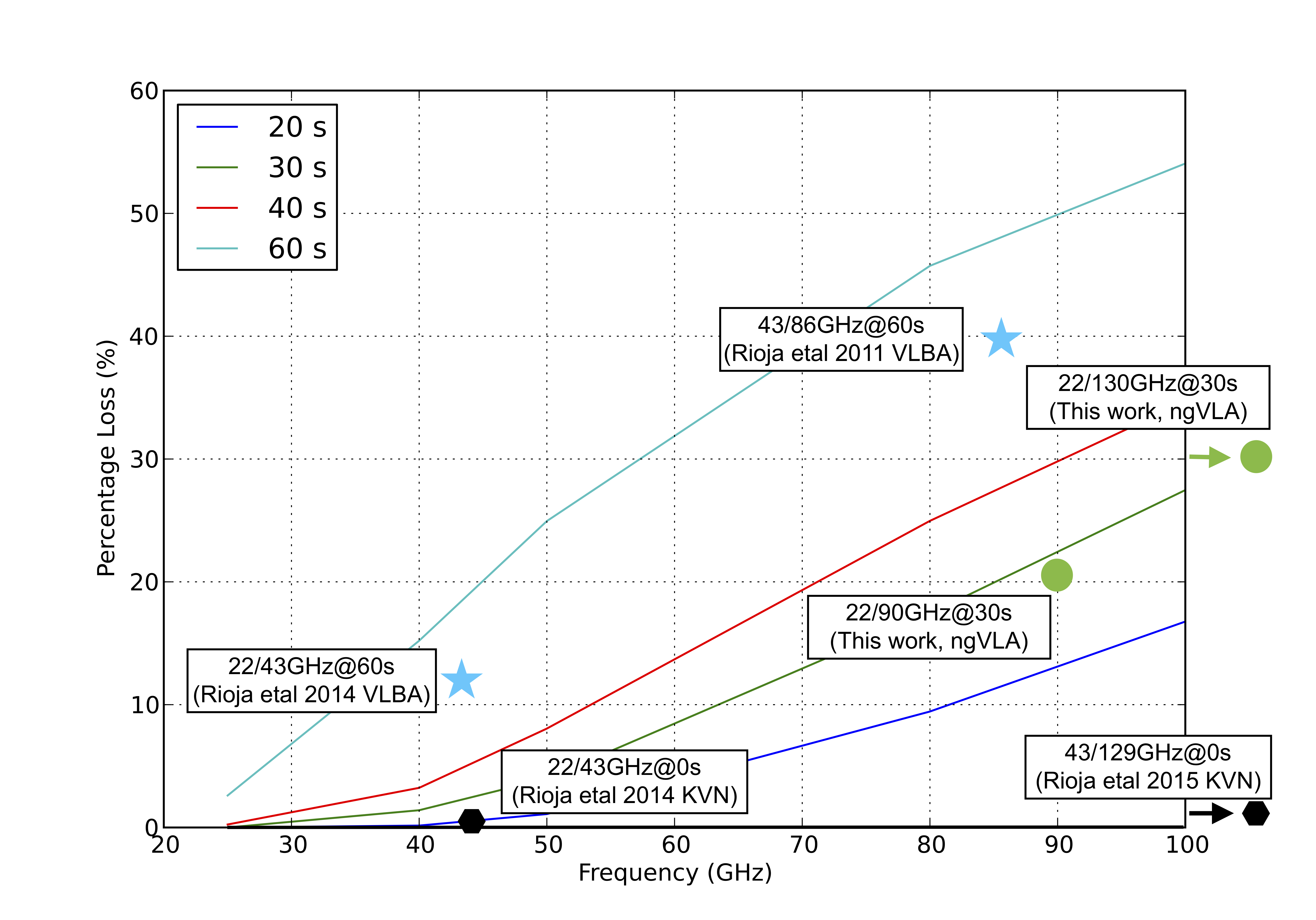}
    \caption{Predicted phase connection losses in multi-frequency calibration, as a function of frequency and of switching duty cycle with lines for 60s (cyan), 40s (red), 30s (green) and 20s (blue). Simultaneous observations would have zero loss, and is indicated with a black line on the x-axis. 
    Overplotted as symbols are the fractional flux recovery of the images made from the same simulated dataset using SFPR (in green circles) and results from the literature using VLBA (red stars) and KVN (black hexagons) observations \cite{rioja_11a,rioja_14}. Losses are over 10\% at the highest frequencies even for the fastest conceivable duty cycle of 20s, whereas simultaneous observations would have zero losses (in these thermal noise free cases).}
    \label{fig:ngvla}
\end{figure}

In this section we consider the impact on the calibration performance from using frequency-switching compared to the simultaneous multi-frequency capability, for the ngVLA (or any other interferometer). This analysis was undertaken for the `ngVLA Community Studies Program'. 
We have carried out simulation studies, to assess the astrometric impact of the options. We used ARIS \cite{a07}, which has a very complete atmospheric model, to provide realistic atmospheric conditions over a subset of ngVLA stations. 
The simulations comprised of SFPR observations with 16 antennas, with baselines from 27 to 500\,km (spanning the edge of the core and the longest baselines of the proposed array), comprising a range of duty cycles for the frequency-switching between zero (i.e. simultaneous) and up to 60s. 
The array response to point sources was generated, with no thermal errors and nominal atmospheric conditions, at the four frequency bands of the KVN. The lowest frequency (22\,GHz) was used as the reference band for the SFPR analysis.
Our first metric is to use the fractional flux recovery of the SFPR images compared to the input model, which allows a direct measurement of the cumulative random phase errors $\Delta \Phi$ ($e^{-\Delta \Phi^2/2}$ \cite{TMSv2}).
Additionally we considered the effect of phase connection errors, which approximates the first metric.
We derived the loss of phase connection based on comparison between the forward predicted phase (using the rate and phase solutions) and the actual phase, for the range of duty cycles. We define failure as where the error is greater than a half cycle. 
Longer intervals, obviously, more often fail to correctly predict the actual phase.
Figure \ref{fig:ngvla} shows the fraction of incorrectly predicted solutions as lines, up to a target frequency of 100\,GHz. 
Overplotted as symbols on this figure are results from the SFPR imaging of the simulated data and empirical results from the literature. The two metrics agree well with each other, and are in reasonable agreement with the empirical results -- for which the weather conditions were excellent rather than nominal.

The figure shows that, at the highest frequencies, the losses are over 10\%, even for the fastest conceivable duty cycle of 20s; whereas simultaneous observations would have zero losses. 
Note that even at lower frequency, for weaker sources, one would still introduce significant errors, as the duty cycles would have to be longer to overcome thermal limits; here too simultaneous observations would still result in zero interpolation errors.
These studies underline the benefits of simultaneous multi-frequency observations, and support the additional effort required to design systems with this capability, at least in mm-wavelengths.

\section{cm/m-VLBI Microarcsecond astrometry using multi-calibrator methods}

\subsection{Ionosphere and conventional PR}\label{sec:iono_intro}
Strategies to correct for the bulk of the propagation medium effects that work well in the regime dominated by the tropospheric errors, fail at lower frequencies ($\nu\lesssim$8GHz) where the ionosphere is the dominant contribution.
Because the ionosphere has spatial structure, i.e. it has direction dependent effects, the PR calibration derived from a reference source along a different direction will leave residuals on the target.
The standard strategy to overcome this is to use a very close calibrator (e.g. in beam), to minimize the relative residual errors.
%
To provide 10\,\uas\ astrometric accuracy using  PR with a single calibrator source at 1.6\,GHz one would require an extremely nearby reference source,
no more than $\sim \,1$\amin\ away from the target 
\cite{a07};
additionally, precise astrometry requires sufficient signal-to-noise-ratio, that is, it requires both very close and strong sources.
The projected source counts suggest that there will not be sufficient calibrator source density to provide this, even at the sensitivities of SKA baselines \citep{ska_135,ska_vlbi}. Since the calibrator availability is limited, this approach has a limited applicability, hence new methods are required.


\subsection{\MV\ astrometric method and demonstration with existing instruments}

The \MV\ calibration method offers the potential to achieve precision astrometry for many targets at the low frequency regime dominated by ionospheric disturbances.
\MV\ VLBI has been reviewed elsewhere \cite{rioja_17} (see also \cite{atmca} and \cite{doi_06}), and here we only summarize the method and benefits. 
It consists of using observations of (at least) three calibrators surrounding the target, 
and uses a 2D interpolation in the visibility domain to provide corrections of the spatial atmospheric distortions along the direction of the target. 
This effectively results as having a virtual calibrator in the same direction as the target.
The \MV\ method is a follow up from the Cluster-Cluster techniques demonstrated in the 1990s \cite{rioja_02}.

The astrometric utility has been empirically demonstrated using VLBA interleaving observations of 
the target and multiple calibrators with angular separations ranging from 0.4\deg\ to 6\deg, over two epochs separated by a month, at 1.6\,GHz.
All sources were quasars, which allows us to measure empirically the astrometric accuracy and precision, using the repeatability of \MV\ results compared to those of PR.  
The main conclusions of this study reveal that:
{\em i)} the superior astrometric precision of \MV, by more than an order of magnitude compared to conventional PR, even for the closest source (0.4 deg away). 
{\em ii)} \MV\ reduced the systematic errors to the thermal noise limit, 
which highlights the essential role of increased sensitivity with next-generation instruments.

\begin{figure}
    \centering
 \includegraphics[width=0.6\textwidth]{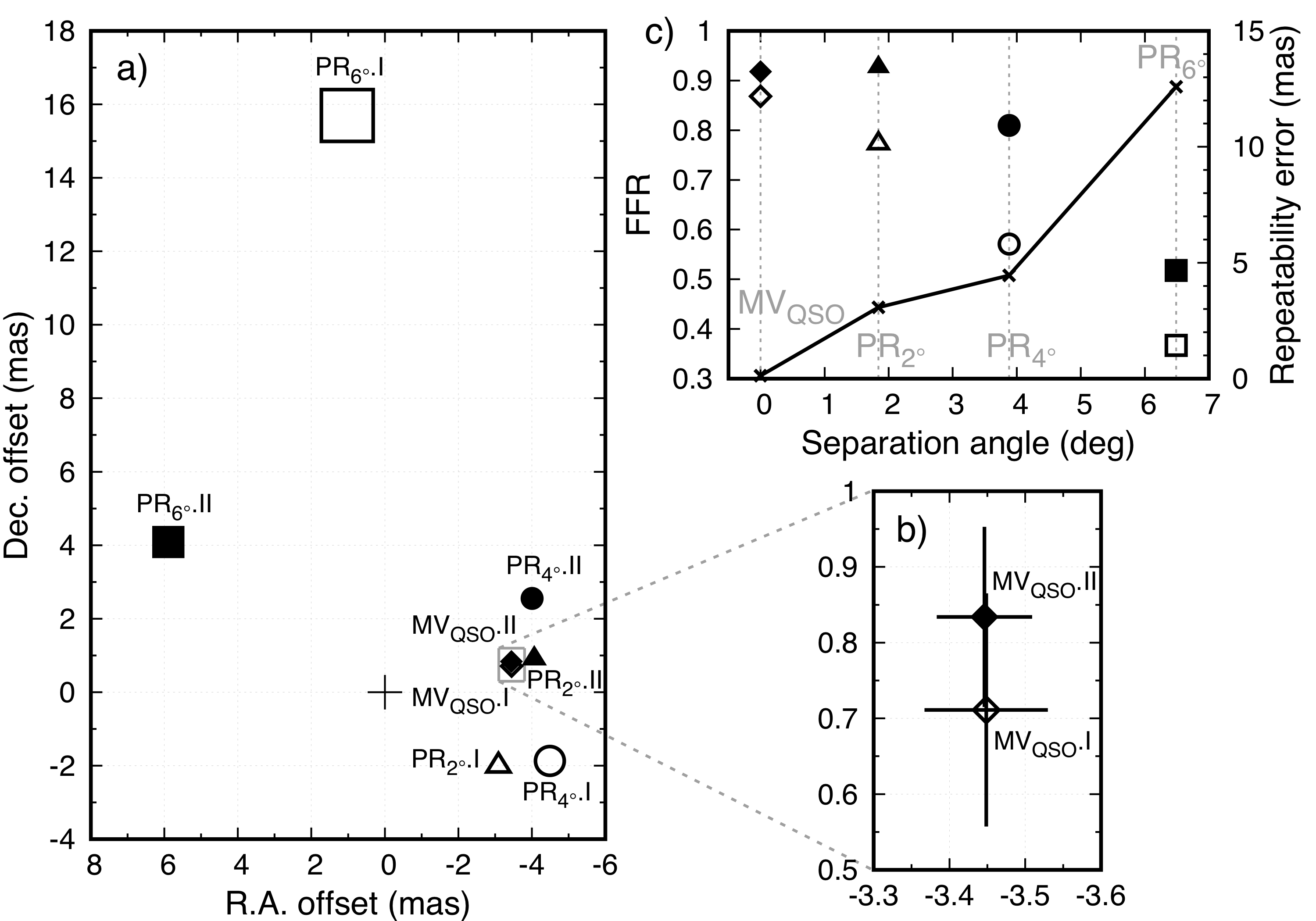}
\caption{ {\it a)} Astrometric offsets in the angular separations
  measured with \MV\,  PR$_{2^o}$,  PR$_{4^o}$  and
  PR$_{6^o}$ analyses (see \cite{rioja_17} for description),
  from the observations of quasars at two
  epochs, with respect to the catalogue position. The size of the plotted symbols
  corresponds to the estimated thermal noise error in each case. The labels describe the calibration methods (MV, PR) and epoch of observations (I, II). {\it b)}  Zoom for \MV\ astrometric solutions. The error bars are the thermal noise errors. Both epochs agree within the error bars.
 {\it c)} Solid line shows the corresponding repeatability astrometric errors versus
 the angular separation between target and calibrator for PR analysis,
 and for an effective $0^o$ separation for \MV. Filled and empty
 symbols show the Flux Fractional Recovery quantity versus angular
 separation for \MV\ (diamond), PR$_{2^o}$ (triangle), PR$_{4^o}$ (circle),
 and PR$_{6^o}$ (square) analyses, for epochs I (empty) and II
 (filled). \label{fig:astro}}
\end{figure}

These findings show that, by using the \MV\ calibration method with three calibrators, one can achieve an order of magnitude improvement in the astrometric accuracy of the target, limited by thermal noise, compared to PR with a single calibrator at similar angular separation. This is in agreement with our simulation studies \cite{rioja_09}.
Even better results, yielding ultra precise astrometry, will come with the next generation instruments, which provide increased sensitivity 
and  simultaneous observations of the multiple sources, to optimize the mitigation of systematic errors.

\subsection{Astrometry in the SKA era}

The SKA provides two orders of magnitude increase in collecting area, and thus sensitivity, over current facilities.  Therefore when cross correlated with existing VLBI facilities there will be an order of magnitude increase in sensitivity, and hence of the astrometric potential.
%
The increased collecting area also results in narrow FoVs, which prevent the simultaneous observations of multiple calibrators to mitigate the systematic errors with \MV.  
Both wide FoVs and high sensitivity are possible for instruments with multibeam technologies.

The development of multibeam capabilities for radio telescopes has undergone a huge leap forward in the last years. 
Currently Phase Array Feeds (PAF) provide a continuous spatial coverage over wide FoVs, that extend by order magnitude over that of a single pixel receiver. 
Other than the advantages for survey speeds, these provide benefits for VLBI astrometry, and are almost mandatory for huge single telescope collecting areas.   
Arrays such as SKA-Mid, and the SKA precursor MeerKAT, will use multiple tied-array beams, steerable along directions within the 15\,m telescope primary beam,
for joint observations with other VLBI telescopes. 
This could be important for deep VLBI surveys of sources close together on the sky, but the major application would be to improve the compensation of the propagation medium effects by using multiple calibrators and enabling ultra precise astrometry.

Radio astronomy in general, and VLBI in particular, will get a big boost with the massive FAST telescope (China, 500m), the world's most sensitive radio telescope.
FAST is a SKA pathfinder with a collecting area similar to SKA, which makes it highly desirable for joint VLBI astrometric observations.
The narrow FoV and limited pointing capabilities complicates the observations of multiple sources. 
Nevertheless it is currently equipped with a 19-beam receiver, which could enable precise astrometry using \MV.
This effectively extends the narrow FoV to that of a 30 m single pixel telescope. There are plans to install a PAF, that will result in an even better astrometric performance. 


With an increasingly large number of telescope with multibeam capabilities the perspectives for low frequency VLBI are most promising.
The largest telescopes in Europe are equipped with PAFs: Effelsberg, Jodrell Bank, SRT, and WSRT (with tied-array beams).
There are many global VLBI stations that currently have or are developing plans to install these: ASKAP, Parkes, Green Bank, FAST, and Arecibo.

Ionospheric studies, the dominant source of errors, are fundamental for our preparation for  SKA-VLBI. Dodson
\cite[these proc.]{dodson_evn_18} presents results from a study of the spatial structure of ionospheric fluctuations, using observations with the MWA -- the SKA-Low pathfinder. 
This concludes that \MV\ SKA-VLBI would be able to achieve the systematic error levels required for ultra precise astrometry ($\sim$1$\mu$as), matching the thermal limits. This study informs the design requirements for a minimum number of six tied-array beams for SKA.

\section{Review of recent advances in astrophysical applications} 

Astrometry continues to demonstrate its wide applicability and we present a limited review of the current state-of-the-art astrometrical results, highlighting aspects of calibration methods, since the last EVN symposium. 

\subsection{Evolved Stars Studies: Record frequency of bona-fide maser astrometry}

The KVN is carrying out a systematic SFPR astrometry study of a sample of 16 evolved stars as a Key Science Program (KSP) \cite[these proc.]{poster_cho_18}, with a factor of three increase in the highest frequency ever measured.
First results from these simultaneous multi-epoch astrometry observations of water maser emission at 22\,GHz and SiO maser emission at 43.1/42.8/86.2/129.3\,GHz, are starting to be released (e.g.  VX Sagittarii\cite{yoon_18}, R Crateris\cite{kim_18} \cite[these proc.]{yun_evn_18,poster_cho_18,poster_yoon_18}). 
The outcomes are bona-fide astrometrically registered images of the spatial and temporal distribution of the emission at all transitions, free of any astrometric assumptions. 
The KSP will provide new insights into this crucial stage in stellar evolution, 
illuminating the mass loss processes and the development of asymmetry in CircumStellar Envelopes with stellar pulsation. Thus providing a wealth of information to discriminate between the competing proposed mechanisms, of radiative or collisional pumping for stellar masers.

\subsection{Galactic Structure: Record parallax distance and 6.7\,GHz astrometry}

VLBI parallax measurements of distances to hundreds of H$_2$O maser sources are revealing for the first time the spiral arms structure of the Milky Way.
Combined with the proper motion this data can be modelled to yield the fundamental parameters of the Galaxy \cite{reid_14}.
%
A recent result from the BeSSeL program is a record breaking distance determination for G007.47+00.05, measuring a parallax of 49$\pm$6$\mu$as, or a distance of $20.4^{+2.8}_{-2.2}$kpc \cite{sanna_17}, made with the VLBA.
Hitherto distances to the far side of the Milky Way have been impossible to measure accurately, because the parallax is very small, making it challenging for VLBI, and because the interstellar dust blocks optical light from those regions, making it also impossible for {\sc gaia}. 

Additionally the radio programs are starting to derive precise astrometric results from observations of 6.7\,GHz methanol masers (\cite{xu_16,wu_19} and \cite[these proc.]{rygl_evn_18}), using an image-based variation of \MV\ \cite{reid_17}, a strategy to minimize the errors on parallax estimates, which has been very successful. 

\subsection{Recollimation shocks in AGNs: mm-VLBI Astrometry without a calibrator source} 

MFPR is a multi-frequency calibration astrometric method with a unique feature: it does not require observations of another source, other than the target. 
It is particularly useful in cases when calibrator availability is an issue. 
It is a variation of SFPR, which uses the observations of the target source at a wider range of frequencies to mitigate the errors introduced by the atmospheric (troposphere and ionospheric) propagation. 
The MFPR method is described in \cite{dodson_17}, along with a demonstration using VLBA astrometric observations of BL\,Lac from 1.3 to 86\,GHz, to reveal the first tentative empirical detection of recollimation shocks in AGNs. 
This project drove the development of MFPR; it could not have been done in any other way, due to the absence of suitable calibrators even within the wide angular separations of many degrees suitable with SFPR. This is an example of how the development of new calibration methods widen the applicability of astrometry to address new astrophysical problems.

\subsection{AGN core-shift studies: Wide angle astrometry at 43\,GHz}

Abell\'an \cite{abellan_18} presented a successful demonstration of precise astrometric measurements with the University of Valencia Precision Astrometry Package (UV\,PAP), of the angular separations between several AGNs that are up to 15\deg\ apart, at 43\,GHz, using VLBA observations. 
This is achievable because of their parametric approach, where every individual contribution to the observables is precisely solved for, 
which allows the authors to overcome the normally tight constraints on the angular separation between sources using PR.
UV PAP global differential phase-delay astrometry analysis relies on the possibility of simultaneous fitting of all the parameters that account for the
geometric, the instrumental and propagation medium contributions to the observables, together with the source position.
Previous UV PAP demonstrations were at lower frequencies. With increasing challenges for higher observing frequencies, 43\,GHz is the
highest frequency demonstrated so far, probably reaching the limit of the applicability of this approach.
%
The precision astrometry measurements at 15 and 43\,GHz were compared, to study the jet physical conditions in a complete sample of 13 radio-loud AGNs around the North Celestial Pole,  using the ``core-shifts'' measurements, and to compare astrometric methods, among them with the multi-frequency SFPR calibration method. 
Their ``core-shift'' values agree with predictions from opacity and synchrotron self-absorption effects \cite{bk_79}.

\subsection{PSR$\pi$: in-beam PR astrometry at low frequencies} 

The major 1.4\,GHz VLBA pulsar astrometric program PSR$\pi$ has recently released the `data-paper' \cite{deller_18} for the project.
PSR$\pi$ is a beautiful demonstration of a successful in-beam PR experiment. 
Out of the 110 initial pulsar list an in-beam reference source could be detected for 60 targets. The minimum pulsar-reference source separation was only 0.8\amin\ and the median angular separation was 14\amin.
The median pulsar parallax error is 45$\mu$as, from typically nine epochs of observations. 
It is worth noting that the errors in the case with smallest angular separation are exactly those of the median separations, because the astrometric errors for that close-but-weak source are dominated by the thermal errors. 
The error analysis in \cite{deller_18} (Eqs. 1 and 2) reaches the same conclusions as described in Section \ref{sec:iono_intro} above; that to achieve systematic errors at the level of 10\,$\mu$as per epoch of observation the angular separation should be $\sim$1\amin, and that the sources should be strong enough so that thermal errors do not dominate. 
This underlines the strength of \MV, for which the effective angular separation can be essentially zero using strong reference sources.


\subsection{Radio astrometry in the {\sc gaia}-era}

With the publication of {\sc gaia} DR2 \cite{gaia_dr2}, the comparison between {\sc gaia} parallaxes, and proper motions, and accurate VLBI astrometric measurements for sources that have 
both optical and radio measurements has become a very active field of research. The multiple studies focus on different objects, among them, AGN positions \cite[and references therein]{petrov_18}, also \cite[these proc.]{petrov_evn_18},  star forming regions \cite{kounkel_18} and evolved stars \cite[these proc.]{langevelde_evn_18}.    
One of the interests of these comparisons is to verify {\sc gaia} results.  Among them, the parallax values for the Pleiades cluster is particularly important because of their role  in the astronomical distance ladder.
Previous comparisons of VLBI and Hipparcos parallaxes resolved the distance controversy and ruled out the Hipparcos value \cite{melis_14}.
The new {\sc gaia} results agree with all VLBI Pleiades parallax measurements of eight stars that show compact emission \cite{MelisNew}. 
%
VLBI radio astrometry and {\sc gaia} astrometry are complementary in many aspects; for example, dust absorption prevents {\sc gaia} observations through the Galactic plane or of deeply embedded young stars, whereas VLBI will not be able to match the sheer number of {\sc gaia} measurements.

\section{Ongoing multi-frequency technological developments relevant to astrometry}

Astrometry has benefited regularly from technological developments that have allowed leaps in performance.
Two ongoing developments are providing the required technological solutions to get the best from the new multi-frequency methods described above, for astrometry. 
%

In the mm-wavelengths, the Compact Triple-band Receiver (CTR) \cite{han_17}, \cite[these proc.]{han_evn_18} is a new implementation of the KVN multi-frequency receivers, with 
three frequency bands, K (18-26GHz), Q (35-50GHz) and W (85-115GHz)) fitted into a single cryostat.
The reduction in size makes this technical solution very versatile for installation in telescopes where space availability is an issue, as is usually the case.
It shows great promise to provide mm-wave Multi-Frequency capability to a wide number of observatories, greatly enhancing the VLBI array suitable for SFPR. 

In the cm-wavelengths, BRoad bAND (BRAND-EVN)  \cite[these proc.]{brand} is a project to build a proto-type prime-focus receiver for the EVN (and other telescopes), with an innovative very wide bandwidth that covers 1.5 to 15.5\,GHz. 
Achieving its full sensitivity potential requires coherent fringe-fitting over the full band, which in turn preserves the chromatic astrometry information between the simultaneous images at different frequencies (in a similar fashion as described above in mm-waves).
The RadioNet project RINGS, for the analysis of such wideband data, is developing a simultaneous solution for both atmospheric contributions (that follow $\nu$ and $\nu^{-1}$ respectively) as well as the intrinsic chromatic effects from the sources themselves.

\section{Conclusions}

There are exciting developments in the field of astrometry, with new technologies and new methods rapidly advancing the options and capabilities. 
%
%
Existing facilities are adding new receivers that provide wider bandwidths and FoVs. 
The next-generation of radio telescopes will provide massive increases in sensitivity, with the potential for ultra-high astrometric precision.
%
However astrometry already rapidly reaches the systematic limits and increased sensitivity alone will not improve accuracy.
Achieving the science goals of the next-generation observatories requires a parallel development of advanced calibration methods along with the technological solutions.
These new methods are under development on multiple fronts: SFPR, and related methods, relevant for mm-VLBI with ngVLA, \MV\ for m/cm-VLBI with SKA and pathfinders, and wide-angle astrometry with UV\,PAP. 
Technologies such as PAFs, for wide FoVs, and CTR for multi-frequency capabilities, are exciting examples of new hardware developments. 
The simultaneous development of technologies and methods will bring us in hitherto unimagined levels of precision in astrometry.

\section*{Acknowledgements}
This presentation has received  funding from the European  Union's Horizon 2020 
research and innovation program under grant agreement No 730562 [RadioNet]

\begin{footnotesize}
\bibliographystyle{JHEP}
\bibliography{sfpr}  
\end{footnotesize}

\end{document}